\documentclass{article}

\usepackage[utf8]{inputenc}
\usepackage{amsfonts}
\usepackage{amsthm}
\usepackage{graphicx}
\usepackage{amsmath}
\usepackage{subcaption}
\usepackage{soul}
\usepackage{hyperref}
\usepackage{authblk}

\newcommand{\R}{\mathbb{R}}

\newtheorem{df}[equation]{Definition}

\newtheorem{ex}[equation]{Example}

\title{Musical Stylistic Analysis: A Study of Intervallic Transition Graphs via Persistent Homology}

\author{Mart\'in Mijangos}
\author{Alessandro Bravetti}
\author{Pablo Padilla}
\affil{Instituto de Investigaciones en Matem\'aticas Aplicadas y en Sistemas, 
Universidad Nacional Aut\'onoma de M\'exico, A.~P.~70543, M\'exico, DF 04510, Mexico}
\begin{document}
\maketitle
    \begin{abstract} 
Topological data analysis has been recently applied to investigate stylistic signatures 
and trends in musical compositions. A useful tool in this area is Persistent Homology. 
In this paper, we develop a novel method to represent a weighted directed graph as a  finite metric space and then use persistent homology to extract useful features. We apply this method to weighted directed graphs obtained from pitch transitions information of a given musical fragment and
use these techniques to the study of stylistic trends. 
In particular, we are interested in using these tools to make quantitative stylistic comparisons. 
As a first illustration, we analyze a selection of string quartets by Haydn, Mozart and 
Beethoven and discuss possible implications of our results in terms of different approaches by these 
composers to stylistic exploration and variety. We observe that Haydn is stylistically the most conservative, 
followed by Mozart, while Beethoven is the most
innovative, expanding and modifying the string quartet as a musical form.
Finally we also compare the variability of different genres, namely minuets, allegros, prestos and adagios, 
by a given composer and conclude that the minuet is the most stable form of the string quartet movements. 
    \end{abstract}

    \section{Introduction}
Several geometrical and topological representations of music have been proposed 
over the past decades in order to extract both qualitative and
quantitative features from musical works and make precise comparisons between works by the same author, 
by different authors, among genres, etc. 
(see e.g. G. Mazzola's ``The Topos of Music" \cite{mazzola2012topos} as well as several papers in the  
more recent book ``Computational Music Analysis" edited by D. Meredith~\cite{meredith2016computational}).
In particular, once a piece of music has been represented as a topological object, we have at our disposal powerful methodological tools.
Recently, topological data analysis, TDA, has been applied to the study of musical works, in particular Persistent Homology, a popular tool of TDA.  
For example in \cite{bergomi2015dynamical} and \cite{bergomi2020homological} persistent homology 
is applied to deformations of the \textit{Tonnetz} that vary over time. 
In~\cite{liu2016applying} persistent homology is incorporated in a convolutional neural network  for an automatic music tagging.

Once a 
topological representation in the form of a simplicial complex of a piece of music has been chosen, TDA can be applied. 
In the present paper, we propose first to construct a directed graph associated 
to a piece and then a novel way of obtaining a simplicial complex from it. 
Once this is done, we compute persistent homology and summarize the results as barcodes. 
We then use statistical and information-theoretical quantities to analyze them.
After developing the tools described above, we apply them to study stylistic features in a specific corpus, namely, 
a sample of the slow movements from string quartets by Haydn, Mozart and Beethoven. 
We propose that the results obtained with our methods can be effectively used to support musicological claims about 
the different attitudes these three composers had in terms of the exploration of stylistic possibilities, 
confirming the traditional view of Haydn as the initiator of the form and Beethoven as a natural innovator.

This paper is organized as follows: in Section \ref{background} we review some basic concepts from algebraic topology and persistent homology. 
We also state the definitions of graph theory that we will use. 
In Section~\ref{graphs} we introduce a novel method to obtain information of weighted directed graphs by means of persistent homology.
In Section~\ref{analysis} we first describe how to obtain a weighted directed graph from a musical work and then apply to 
it the techniques described 
in Section~\ref{graphs}. 
We propose statistical and information-theoretical quantities to extract features from the data. 
In Section~\ref{minuets}  
we reach conclusions in two directions. 
One about stylistic differences between slow movements from three different authors and 
the other about stylistic differences 
between various types of composition of a given author. 
We end the paper with conclusions and further research perspectives in Section~\ref{conclusions}. 
We remark that all codes developed for this paper are available at 
\href{https://github.com/MartinMij/TDA-SQ}{https://github.com/MartinMij/TDA-SQ}.

\section{Background}\label{background}

\subsection{Simplicial complexes and homology}
 In this subsection we review some basic algebraic topology concepts. 
 More details can be found in Munkres~\cite{munkres2018elements} and Hatcher~\cite{hatcher2005algebraic}.

Intuitively, a simplicial complex is a topological space composed of smaller pieces called simplices. 
These simplices are points, lines, triangles, tetrahedra and their higher dimensional generalizations. 
A formal definition of a simplicial complex is the following.
\begin{df}\rm
An (abstract) \textit{simplicial complex} is a set $K$ of finite subsets of a set $S$ in such a way that $\{v\}\in K$ for all $v\in S$
and if $\tau\subseteq\sigma\in K$ then $\tau\in K$. Elements $\{v\}\in K$ are the \textit{vertices} of 
$K$ and $S$ is its \textit{vertex set}. Any element in $K$ is a \textit{simplex} of $K$. 
A \textit{face} of $\sigma\in K$ is a simplex $\tau$ such that $\tau\subseteq\sigma$. 
We say that the \textit{dimension of a simplex} $\sigma$ is $p$ or that $\sigma$ is a $p$-simplex if 
$|\sigma|=p+1$ where $|\sigma|$ denotes the cardinality of the set $\sigma$.  
The \textit{dimension of the complex} $K$ is defined as the largest dimension of any of its 
simplices or as infinite if there is no upper bound in such dimensions.  
%\pabloscomment{What is $||$?\checkmark}\alescomment{can $S$ be infinite?\checkmark Yes, I've added remarks}
\end{df}
\begin{df}\rm
Given a simplex $\sigma$, we order its vertices as $(v_0, v_1,...,v_p)$ and define an equivalence relation where 
$(v_0,...,v_p)\sim (u_0,..., u_p)$ if there exists an even permutation $s\in S_p$ such that $(v_0,...,v_p)=(u_{s(0)},...,u_{s(p)})$. 
We define an \textit{oriented simplex} $\sigma$ as an equivalence class of this relation and we denote it as  $[\sigma]=[v_0,v_1,...,v_p]$. 
\end{df}

We will consider that $[\sigma]=-[\tau]$ if $\sigma=\tau$ and $[\sigma]\neq [\tau]$. 
From now on, simplices are taken oriented and we will just write $\sigma$ for $[\sigma]$.%\alescomment{A different orientation corresponds to an odd permutation?\checkmark Yes, I've changed the words}
%\alescomment{``simplices are taken ordered'', means ``simplices are taken oriented''?\checkmark Yes}

\begin{df}\rm
A \textit{subcomplex} $L$ of $K$ is a subset of $K$ that is a simplicial complex itself. In particular, 
for a given $i\in\mathbb{N}$, the $i$\textit{-skeleton} of $K$ is the subcomplex composed by all the simplices of $K$ 
of dimension at most $i$.
\end{df}

\begin{df}\rm
Given two simplicial complexes $K$ and $L$, a \textit{simplicial map} $f:K\longrightarrow L$ 
is a function that takes the vertices of $K$ into the vertices of $L$ and such that 
if $\sigma$ is a simplex of $K$ then $f(\sigma)$ is a simplex of $L$.
\end{df}
%\begin{df}\rm
%A $p$-simplex $\sigma$ is the convex hull of $p+1$ affinely idependent points $v_0, v_1,..., v_p$ in $\mathbb{R}^d$, that is, points such that $v_1-v_0, v_2-v_0,...,v_p-v_0$ are lineary independent. Explicitly,  
%$$\sigma=\{\lambda_0 v_0+\lambda_1 v_1+\cdots +\lambda_p v_p|\lambda_0, \lambda _1,...,\lambda_p\in [0,1]  \mbox{ and } \sum_{i=0}^p\lambda_i=1\}$$
%We will denote such a simplex $\sigma$ as $\langle v_0,...,v_p\rangle$.
%\end{df}
%We will call vertices of $\sigma$ to the points $v_0, v_1,...,v_p$. As any subset of the vertex set is affinely independent too, its convex hull is a simplex named \textit{face} of $\sigma$. 

%\begin{df}\rm
%A simplicial complex $K$ in $\mathbb{R}^d$ is a set of simplices in $\mathbb{R}^d$ such that
%\begin{itemize}
%    \item if $\sigma\in K$ and $\tau$ is a face of $\sigma$, then $\tau\in K$ and
%    \item for any two simplices $\sigma_1, \sigma_2\in K$ with non-empty intersection, %$\sigma_1\cap\sigma_2$ is a face of both $\sigma_1$ and  $\sigma_2$.
%\end{itemize}
%\end{df}

An example of a simplicial complex which will be useful later is the Vietoris-Rips complex. 
\begin{ex}\rm
Let $(X, d)$ be a finite metric space and $r$ a non-negative real number. 
A \textit{Vietoris-Rips complex} $VR(X, r)$ is a simplicial complex with vertex set $X$ 
and such that a $p$-simplex $[x_0, x_1,...,x_p]$ is in $VR(X, r)$ if $d(x_i, x_j)\leq r$ for all $i,j\in\{0,1,...,p\}$, $i\neq j$, 
or equivalently if $B(x_i, r/2)\cap B(x_j, r/2)\neq \emptyset$, 
where $B(x_i, r/2)$ represents the ball with center $x_i$ and radius $r/2$. 
Observe that if $0\leq r_1\leq r_2$, then $VR(X, r_1)$ is a subcomplex 
of $VR(X, r_2)$.
\end{ex}

\begin{figure}[h]
    \centering
    \includegraphics[scale=0.35]{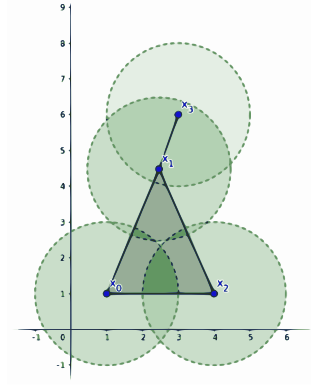}
    \caption{Vietoris-Rips complex $VR(X,3)$ for $X=\{x_0,x_1,x_2,x_3\}$ 
    and where the distance $d$ has
    been chosen to coincide with the corresponding Euclidean distance in $\R^2$.  
    $VR(X,3)$ consists of four 0-simplices, four 1-simplices and one 2-simplex.}
    \label{fig:vr-ex}
\end{figure}

In Fig.~\ref{fig:vr-ex} we provide an illustration of a Vietoris-Rips complex.

We now define homology for simplicial complexes. 

\begin{df}\rm
Let $K$ be a simplicial complex and $F$ a field. 
We indicate by $C_p(K,F)$ 
the $F$-vector space with basis the $p$-simplices of $K$ 
(when it is clear the field over which we are working, we will just write $C_p(K)$). 
An element in $C_p(K)$ is called \emph{a $p$-chain}. 
\end{df}

Note that by definition a $p$-chain is a finite sum of the form 
$$ a_1\sigma_1+a_2\sigma_2+\cdots+a_n\sigma_n\,,$$ 
with $a_i\in F$ and $\sigma_i$ a $p$-simplex. 

\begin{df}\rm
We define \emph{a boundary map} as follows:
    \begin{enumerate}
        \item[i)] for any $p\geq 1$ 
            $$\partial_p:C_p(K)\longrightarrow C_{p-1}(K)$$ 
                is the map that acts on any basis element 
                $\sigma=[v_0, v_1, ...,v_n]$ 
                as 
            $$\partial_p(\sigma)=\sum_{i=0}^p(-1)^i[ v_0,...,\hat{v}_i,...,v_p]$$ 
            where the hat above a vertex indicates that the corresponding vertex has been removed 
        \item[ii)] for any $p<0$, $C_p(K)$ and $\partial_p$ are defined as zero
        \item[iii)] $\partial_0:C_0(K)\longrightarrow 0$ is the zero map.
    \end{enumerate}
$(C_p(K), \partial_p)$, $p\in\mathbb{Z}$, is called a \emph{chain complex}.
\end{df}

It is a basic and most important fact that $\partial_p\circ\partial_{p+1}=0$ 
for all $p\in\mathbb{Z}$. 
We now define the homology groups of the simplicial complex $K$ as follows.

\begin{df} \rm\label{homology}
 For a $p\in\mathbb{Z}$,
 $$H_p(K)=\ker{\partial_p}/\mbox{im}\,\partial_{p+1}\,,$$
 is the \emph{$p$-th homology group of $K$}.
 Any element $c\in H_p(K)$ is called a \emph{homology class}.
 \end{df}
 Note that all homology groups are trivial for $p<0$.
 Moreover,
 they are all vector spaces, as they are quotients of vector spaces over $F$. 
 Hence we can define the \textit{Betti numbers} $\beta_p(K)$ as the dimension
 of the vector space $H_p(K)$. 
  %% % %%In Definition \ref{homology} we have taken the coefficients over a field but instead, 
 %we could have taken any ring (indeed any abelian group). 
 %When coefficients are taken over $\mathbb{Z}$, 

 Intuitively, $\beta_p(K)$ is the numbers of $p$-dimensional holes in $K$ (Hatcher~\cite{hatcher2005algebraic}). 
 For instance, $\beta_0(K)$ represents the number of connected components of $K$, 
 $\beta_1(K)$ is the number of loops and $\beta_2(K)$ is the number of voids or cavities of $K$. 
 %If homology groups over $\mathbb{\Z}$ are torsion free, then the Betti numbers are independent of the coefficient 
 %field as a consequence of the Universal Coefficient Theorem (\cite{hatcher2005algebraic} \$3.A). 
 %\alescomment{In all this part we need references \checkmark}
 
 Note however that in general the Betti numbers depend on the field $F$.
 In order to simplify computations, we will be mainly interested in taking $F=\mathbb{Z}_2$ as the coefficient field. 
 In this case the coefficients are 0 or 1, which allows us to forget orientations of simplices for example. 
 
 As a final remark, we recall that, given a simplicial map $f:K\longrightarrow L $, 
  for all $p\in \mathbb{Z}$ we have the
 \emph{induced maps} 
 $$(f_*)_p:H_p(K)\longrightarrow H_p(L)$$ 
 given by
 \begin{align}\label{functoriality}
 \begin{split}
     (f_*)_p:%H_p(K)&\longrightarrow H_p(L)\\
     \bar{c}\longmapsto\overline{f(c)}
 \end{split}
 \end{align}

\subsection{Persistent homology}
When dealing with data analysis we typically want a way to assess the importance of different aspects.  
For instance, we may want to identify noise from real features, 
or we may want to pay special attention to some features within a certain threshold. 
This is what persistent homology is useful for:
homology alone captures the features of a `static' simplicial complex.
Then the key idea is to take simplicial complexes at different scales to  
represent our data, and then assess the importance of different features based on how long they persist when we move 
along all scales.
We make this idea precise as follows.
\begin{df}\rm
A \textit{filtration} of  a simplicial complex $K$ is a sequence of nested simplicial 
complexes $K_0\subseteq K_1\subseteq\cdots\subseteq K_n\subseteq K$.
Moreover, we set $K_i=\emptyset$ for $i<0$ and $K_i=K$ for $i>n$.
\end{df}

\begin{ex}\label{ex VR}\rm
Given a finite metric space $(X, d)$ and any sequence of non-negative real numbers 
$\epsilon_0\leq \epsilon_1\leq\cdots\leq\epsilon_n$, we have a filtration of Vietoris-Rips complexes 
$$VR(X, \epsilon_0)\subseteq VR(X, \epsilon_1)\subseteq\cdots\subseteq VR(X, \epsilon_n).$$
In this work we will consider filtrations of Vietoris-Rips complexes 
with $\epsilon_0=0$ and $\epsilon_n=\max\{d(x_j, x_i)\neq\infty|x_i, x_j\in X\}.$
\end{ex}

As we saw in Equation~\eqref{functoriality}, an inclusion $\rho^{i,j}:K_i\hookrightarrow K_j$, $i\leq j$, induces a map 
$$(\rho^{i,j}_*)_p:H_p(K_i)\longrightarrow H_p(K_j)$$ %\alescomment{now we lost the $*$ with respect to eq. 7\checkmark Yes, I rewrote to avoid notation of (7).}
for all $p\in\mathbb{Z}$. 
To simplify the notation, from now on
we will drop the subscript~$*$ and just denote these maps as $\rho^{i,j}_p$. 
Given a class $c$ in $H_p(K_i)$ (which can be considered as a $p$-dimensional hole) 
we can track its persistence as we move along filtration by means of the maps $\rho^{i,j}_p$
according to the next definition.

\begin{figure}
 \begin{subfigure}{1\textwidth}
  \centering
  \includegraphics[scale=0.25]{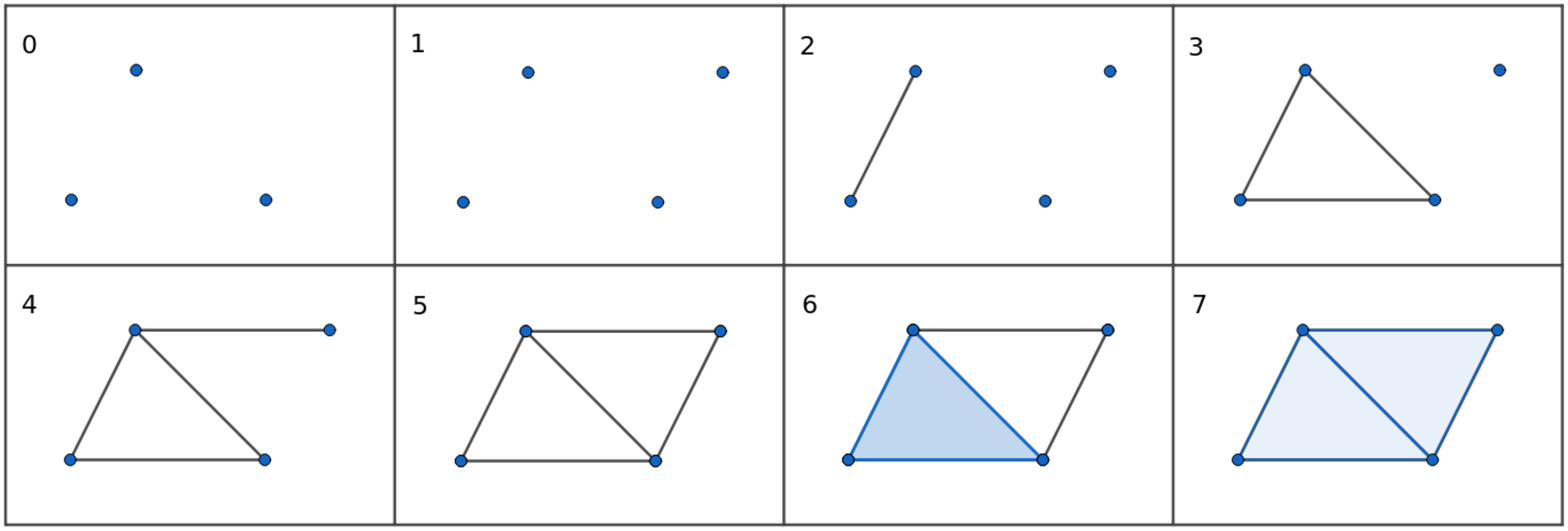}
  %\caption{A subfigure}
  %\label{fig:sub1}
  \end{subfigure}%
  
 \begin{subfigure}{1\textwidth}
  \centering
  \includegraphics[height=5.5cm, width=9cm]{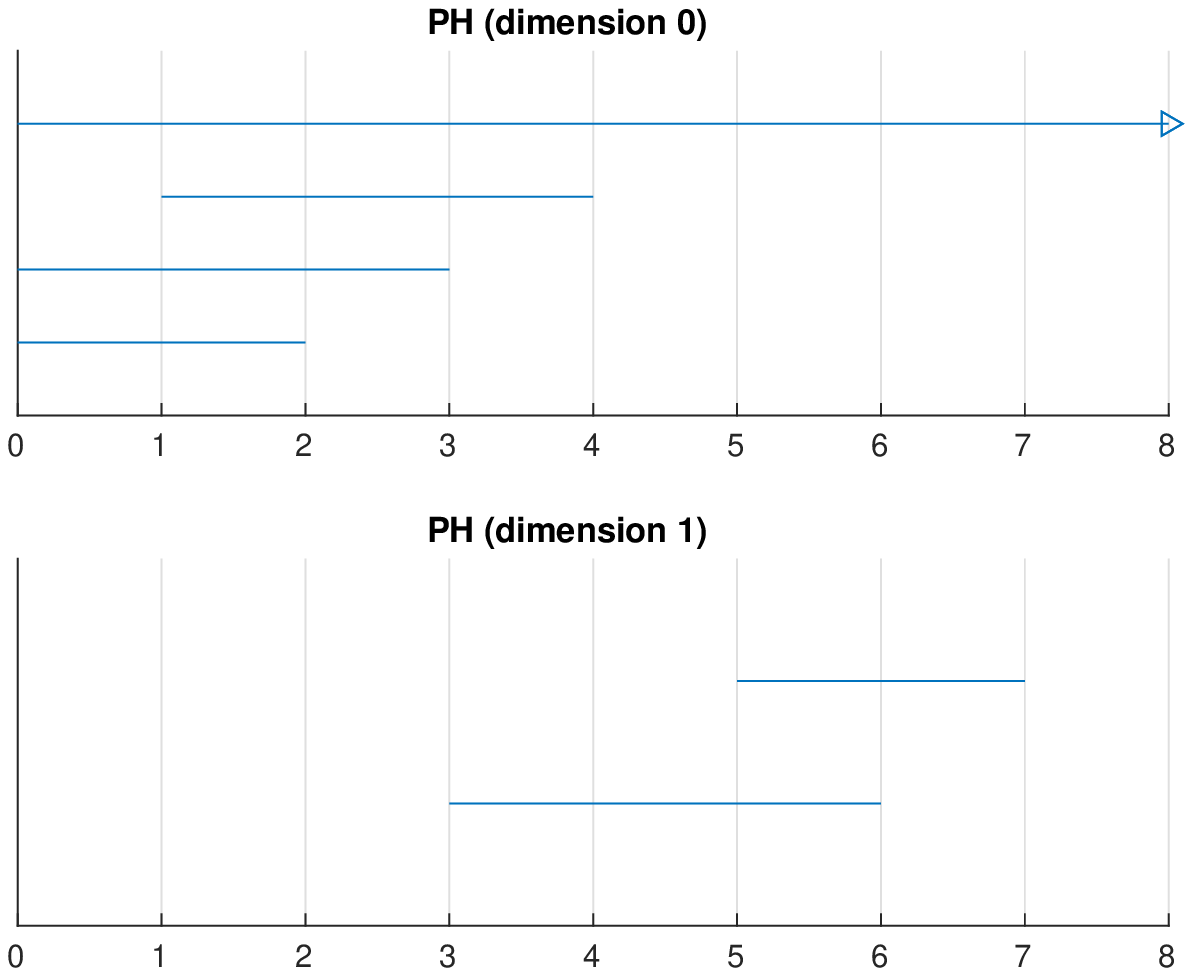}
  %\caption{A subfigure}
  %\label{fig:sub2}
 \end{subfigure}
 \caption{Filtration of simplicial complexes and the corresponding barcodes in dimension 0 and 1. 
    The number in the left-upper corner of each simplicial complex stands for its level in the filtration.}
 \label{fig:ph-ex}
\centering
\end{figure}

\begin{figure}[ht]
    \centering
    \includegraphics[scale=0.32]{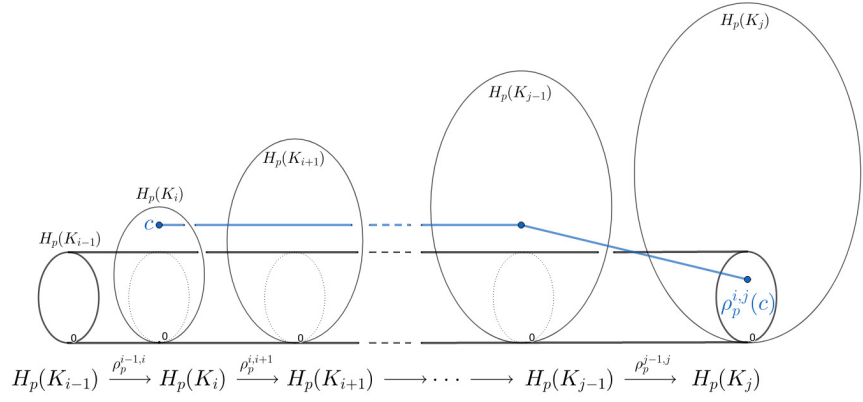}
    \caption{A class $c$ that is born at $K_i$ and dies entering $K_j$.}
    \label{fig:borndie}
\end{figure}

\begin{df}\rm
Given a homology class $c\in H_p(K_i)$ we define its birth and death as~\cite{edels-harer}
\begin{itemize}
\item[i)] We say that $c$ \textit{is born} at $K_i$ if $c\in H_p(K_i)\backslash\{0\}$ and $c\notin \mbox{im}(\rho^{i-1, i}_p).$
\item[ii)] We say that $c$ \textit{dies entering}  $K_j$ 
if $\rho^{i, j-1}_p(c)\notin \mbox{im}(\rho^{i-1,j-1}_p)$ but $\rho^{i, j}_p(c)\in \mbox{im}(\rho^{i-1, j}_p).$
\end{itemize}
Then \emph{the persistence of $c$} is the associated 
half-open interval $[b, d)$ where $b$ represents the level 
in the filtration where $c$ was born and $d$ where it dies. 
If a class never dies we take $d=\infty$.
The multiset given by the intervals of all $p$-homology classes appearing in a filtration $F$ is called 
\textit{the p-persitence barcode}, denoted by $bc_p(F)$. 
\end{df}
In Fig.~\ref{fig:ph-ex} we provide an illustration of a filtration of simplicial complexes, together with 
the corresponding barcodes in dimension 0 and 1,
while in Fig.~\ref{fig:borndie} we present a pictorial representation of a class that is born 
at $K_i$ and dies entering $K_j$.

%\alescomment{I just recalled: if possible all the images in .eps}

\subsection{Directed graphs}
When studying objects with relations among them, 
a natural way of representing the information is through a graph. Now we set the definitions that we will use.  
\begin{df}\rm
An \textit{undirected graph}, or simply a \textit{graph},  is a pair $G=(V, E)$ where $V$ is the set of 
\textit{vertices} and $E$ is a set of 2-element subsets of $V$. 
The elements of $E$ are called the \textit{edges} of $G$. 
\end{df}
Note that this definition does not allow for the existence of edges joining a vertex with itself, 
nor for multiple edges between a pair of vertices. 
Sometimes this graph is also referred to as a \emph{simple graph}. %to avoid confusion.

\begin{df}\rm 
A \textit{directed} graph (or \textit{digraph}) is a pair $G=(V, E)$ where $V$ 
is the set of vertices and $E\subseteq V\times V$ is the set of edges.
\end{df}
Note that this definition allows for the existence of loops, that is, edges joining a vertex with itself. 

\begin{df}\rm
A \textit{weighted} graph (resp.~a weighted digraph) $G^w=(V, E, w)$ is a graph (resp.~digraph) equipped with 
a function {$w:E\longrightarrow\mathbb{R}^+$}, that is, a positive number is associated to each edge. 
Intuitively, such number gives the \emph{weight} of the corresponding edge.
 \end{df}

\begin{df}\rm
Given a weighted graph or digraph $G^w=(V, E, w)$ with finite vertex set $V=\{v_1,...,v_n\}$, 
its \emph{adjacency matrix} is the $n\times n$ matrix $A=(a_{ij})$
whose $(i, j)$ entry is given by
\begin{enumerate}
    \item $a_{ij}=w(v_i, v_j)$, if there is an edge between $v_i$ and $v_j$ 
    \item $a_{ij}=0$, otherwise.
\end{enumerate}
\end{df}

\section{PH of weighted directed graphs}\label{graphs}
Given a directed graph $G$, we would like to apply persistent homology to extract its main features. 
%techniques as persistent homology to obtain features about it. 
To do so we need a filtration of simplicial complexes. 

One way to obtain a filtration is to first 
associate a simplicial complex to a directed or undirected graph by means of the
so-called \emph{neighborhood complex} $N(G)$ or  
the \emph{clique complex} $C(G)$~\cite{Horak_2009}. %\alescomment{Why don't we use these?\checkmark Because we have weighted graphs.}
%Once having a simplicial complex $K$, 
%we can take 
Then a filtration $\emptyset\subset K_0\subset K_1\subset \cdots \subset K_n=K$ %in different ways. 
%For example, 
is given by taking $K_i$ as the $i$-skeleton of $K$.  %\alescomment{What is the $i$-skeleton? \checkmark I updated Definition 3}
%Then, applying persistent homology is possible.
A disadvantage of this approach is that it is hard to adapt it to consider weighted graphs
(see e.g.~\cite{petri2013topological} for an extension to include undirected weighted graphs).
%In~\cite{petri} the authors obtain filtrations of weighted graphs named 
%\textit{Weight Rank Clique filtrations} but these filtrations work only for undirected graphs, and 
%though they affirm that the method can be adapted for directed graphs, 
%they do not give a general way to do it.

Here we propose a new way to obtain a filtration, which is naturally adapted to 
weighted directed graphs.
The key idea is to transform a weighted directed graph into its associated undirected graph (Definition~\ref{def:asswgraph} below),
then use the weights to endow it with
a metric, and finally define the corresponding Vietoris-Rips complexes
to obtain a filtration.

We start by showing how any weighted \emph{undirected} graph 
can be thought of as a finite metric space~\cite{otter2017roadmap}:
    \begin{df}\rm\label{def:graphdistance}
Let $G^w=(V, E, w)$ be an undirected graph with positive weights. 
% and define a distance $d:V\times V\longrightarrow [0,\infty)$ between the vertices as 
%$d(u, v)=1/w$ if $u$ and $v$ are connected by an edge of weight $w$.
For any pair of vertices $u$ and $v$ in $V$ and any path $\gamma(u,v)$ connecting them,
define \emph{the length of $\gamma(u,v)$}, denoted $\ell(\gamma(u,v))$,
as the sum of the inverses of the weights of the edges in $\gamma(u,v)$.
Then define \emph{the distance between $u$ and $v$}, denoted $d(u, v)$, 
as 
    \begin{enumerate}
        \item $d(u,v)=0$, if $u=v$
        \item $d(u,v)=+\infty$, if there is no path $\gamma(u,v)$
        \item $d(u,v)=\min_{\gamma(u,v)}\ell(\gamma(u,v))$, otherwise.
    \end{enumerate}
It is straightforward to verify that $d:V\times V\longrightarrow [0,+\infty]$  
is a (\textit{extended}) metric on $V$.
We call this metric \emph{the weight metric of $G^w$}.
    \end{df}

Note that, in particular, if $u$ and $v$ are joined by an edge of weight $w$, then $d(u,v)=1/w$, 
and that those vertices joined by edges with greater weight are closer in the metric space than 
the ones joined by edges of smaller weight. 
The intuition is that the weight of an edge represents 
the importance of the connection between the two vertices, 
and therefore the more important the connection between the two vertices, the closer they are. 
%closeness of vertices in the metric spaces is equivalent to the importance of their connection.

Clearly the former technique to obtain a metric space from an undirected
graph does not work for \emph{directed} graphs, 
since in this case the so-defined distance would not be a symmetric 
function.
%%%%%%%%Alescomment
%\alescomment{In principle one can bypass this problem by ``symmetrizing'' the distance, i.e.~defining
%$$d_1(a,b):=1/2(d(a,b)+d(b,a))$$
%Why don't we do that? Because we would loss the information of the direction of the edges.}
To overcome this issue, we propose the following strategy:
given a directed graph, we first associate to it an undirected graph 
%\textcolor{red}{which respects as much as possible
%the directional character of the original graph,}
and then apply to this associated graph the above definition
of a distance. Therefore we need the following

\begin{figure}[ht]
\centering
%\begin{subfigure}{.6\textwidth}
%  \centering
  \includegraphics[width=.49\linewidth]{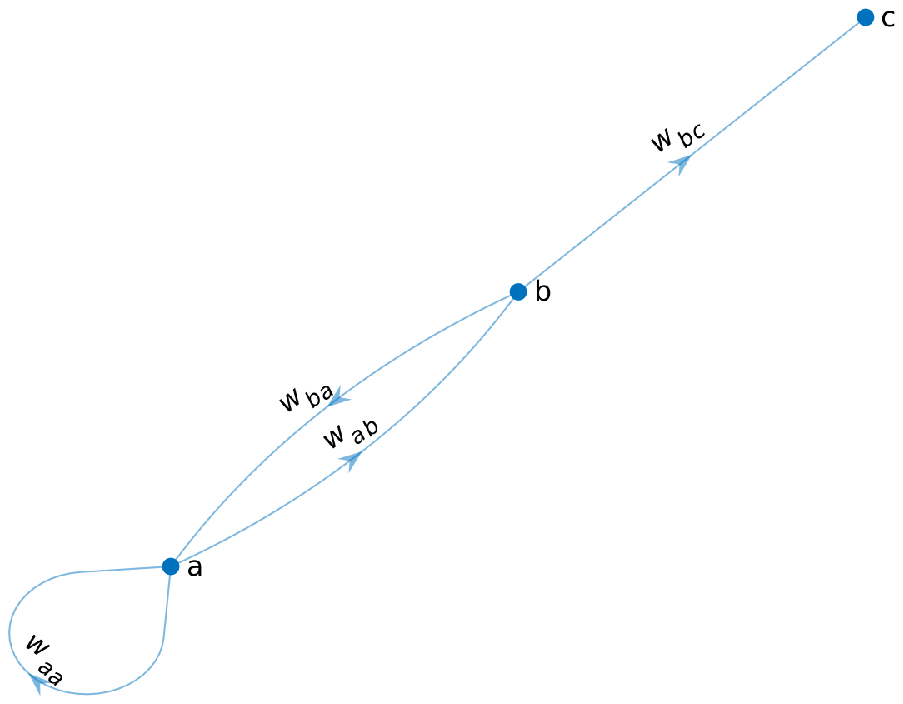}
  %\caption{A subfigure}
  %\label{fig:sub1}
%\end{subfigure}%
%\begin{subfigure}{.6\textwidth}
%  \centering
  \includegraphics[width=.49\linewidth]{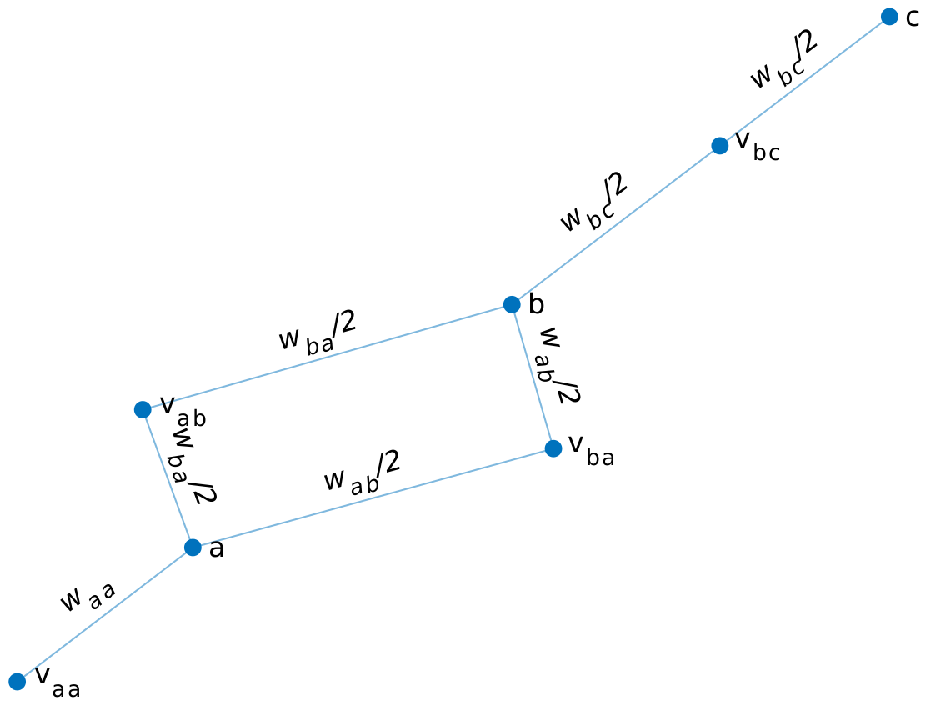}
  %\caption{A subfigure}
  %\label{fig:sub2}
%\end{subfigure}
\caption{A digraph and its associated undirected graph.}
\label{exg}
\end{figure}

    \begin{df}\rm\label{def:asswgraph}
    Let $G=(V, E)$ be a directed graph. We define \emph{its associated undirected graph}
    as 
    the graph 
    $G'=(V', E')$, where
    $$ V':=V\cup \{v_{ab}|(a, b)\in E\}\quad \text{and} \quad
    E'=\{\{a, v_{ab}\}|(a,b)\in E\}\cup \{\{b, v_{ab}\}|(a,b)\in E\}.$$ 
    Moreover,
    if the directed graph $G$ is weighted with weight function $w$, 
    then we define \emph{the associated weight function} $w'$ on $G'$ by 
    $$
    w'(\{a, v_{ab}\})=w'(\{v_{ab},b \})=\frac{1}{2}w(a, b) \quad \text{if}  \quad a\neq b$$ 
    and 
    $$ w'(\{a, v_{aa}\})=w(a,a).$$ 
    \end{df}
The idea behind this construction is to add a vertex $v_{ab}$ for each directed edge $(a,b)$ that works as a label, 
see Figure~\ref{exg}, where the weights $w(u,v)$ are denoted as $w_{uv}$.
Note also the important property that 
if $w$ is a probability distribution over the edges of $G$, then $w'$ is a probability distribution over 
the edges of $G'$.
{This is because} a directed edge in $G$ which is not a loop splits into two edges in $G'$ 
whose weights add up the weight of the original edge
and a loop in $G$ becomes a regular edge in $G'$ with the same weight.

Now that we have an associated weighted undirected graph, the crucial step is completed:
what follows is to apply the construction detailed in
Definition~\ref{def:graphdistance}
to obtain a metric space,
obtain a filtration of Vietoris-Rips complexes {as in Example~\ref{ex VR}}, and
finally use persistent homology to analyze the data. 
%In the next section we will describe the results obtained by app this new procedure for the case 

\section{Stylistic analysis of musical works}\label{analysis}
In this section we show how we can apply persistent homology to obtain features of musical works. Much research has been done trying to
apply classification techniques to music. This ranges from basic statistical tools to sophisticated artificial intelligence methods. The
results often have important musicological and analytical consequences related to authorship, chronology and genre identification. Besides their
theoretical interest, these methodologies might have also practical implications, such as in legal disputes, e.g.~plagiarism issues.
In any case, the question of whether stylistic signatures can be obtained from data analysis has attracted much attention in the past few years.

\subsection{PH of musical works}\label{phofmusicalworks}

In what follows we apply the techniques discussed above to a specific example, namely, 
the stylistic development and identification of the string quartets by Haydn, Mozart and Beethoven. 

First of all, given a musical piece, we need to extract a graph from it.
This is done following the approach recently put forward in~\cite{pablo} and \cite{pablo2022}:
given a score, the main idea is to analyze it by using its distribution of pitch class transitions 
weighted by note durations in seconds (indeed we will take durations modified according to Parncutt's model~\cite{parncutt1994perceptual})
in order to convert it into a weighted directed graph.

To be precise, we have the following

\begin{df}\rm\label{def:musicalgraph}
Let us enumerate the twelve pitch classes C, C\#, D, D\#, E, F, F\#, G, G\#, A, A\#, B from 1 to 12 
respectively, and consider the set of pitch class transitions $X=\{(i, j)|i,j\in\{1,2,...,12\}\}$ where the pair $(i, j)$ 
represents the transition from the pitch class $i$ to the pitch class $j$. In the musical piece, 
a transition $(i, j)$ can appear several times and in each case we consider 
a weighted  duration of the transition as the product of the duration of the tone $i$ and the duration of 
the tone $j$. 
Finally, we 
let $p:X\longrightarrow [0,1]$ be the probability function given by
$$p(i, j)=\frac{\mbox{total weighted durations of transitions }(i, j)}{\mbox{total weighted durations of all transitions}}\,.$$

This information can be organized in a $12\times 12$-matrix $M=(m_{i j})$ where $m_{i j}=p(i,j)$. 
We use a built-in function of the Midi Toolbox~\cite{eerola2004midi} implemented in MATLAB to calculate this matrix.
Crucially, we can now see the matrix $M$ as the adjacency matrix of 
a weighted directed graph whose vertices are the pitch classes 1, 2, ..., 12,
and if $m_{ij}\neq 0$, there is a weighted directed edge from $i$ to $j$  with weight $m_{ij}$. 
So, we have associated a weighted directed graph to the score we started with.
We call this graph the \emph{intervallic transition graph}.
\end{df} 

Now we are in the position to apply the ideas presented in Section~\ref{graphs} to obtain a metric space
and from that we can use 
%Once we have a metric space data, we apply 
persistent homology over the filtration of Vietoris-Rips complexes to obtain the barcodes for each dimension. 

    \begin{ex}\rm\label{ex:barcodesop17H}
    In Figure~\ref{fig:bc} we show the barcodes of dimension 0 and 1 
obtained from the first violin of the second movement of the String Quartet Op.17 by Haydn. 
Note that in dimension 0
there are two infinite bars, which means that there are two connected components in the graph. 
The first component with all vertices appearing in at least one transition, 
and the second component consisting of a single vertex, which in turn represents a  pitch class never played in the work.
We also observe that, as expected, in dimension 0 all the bars in $bc_0(F)$ have zero as their birth time. 
However, this is not true in higher dimensions, as it can be seen clearly in the right panel of Figure~\ref{fig:bc}.
\begin{figure}[ht]
\centering
\begin{subfigure}{.49\textwidth}
  \centering
  \includegraphics[width=1\linewidth]{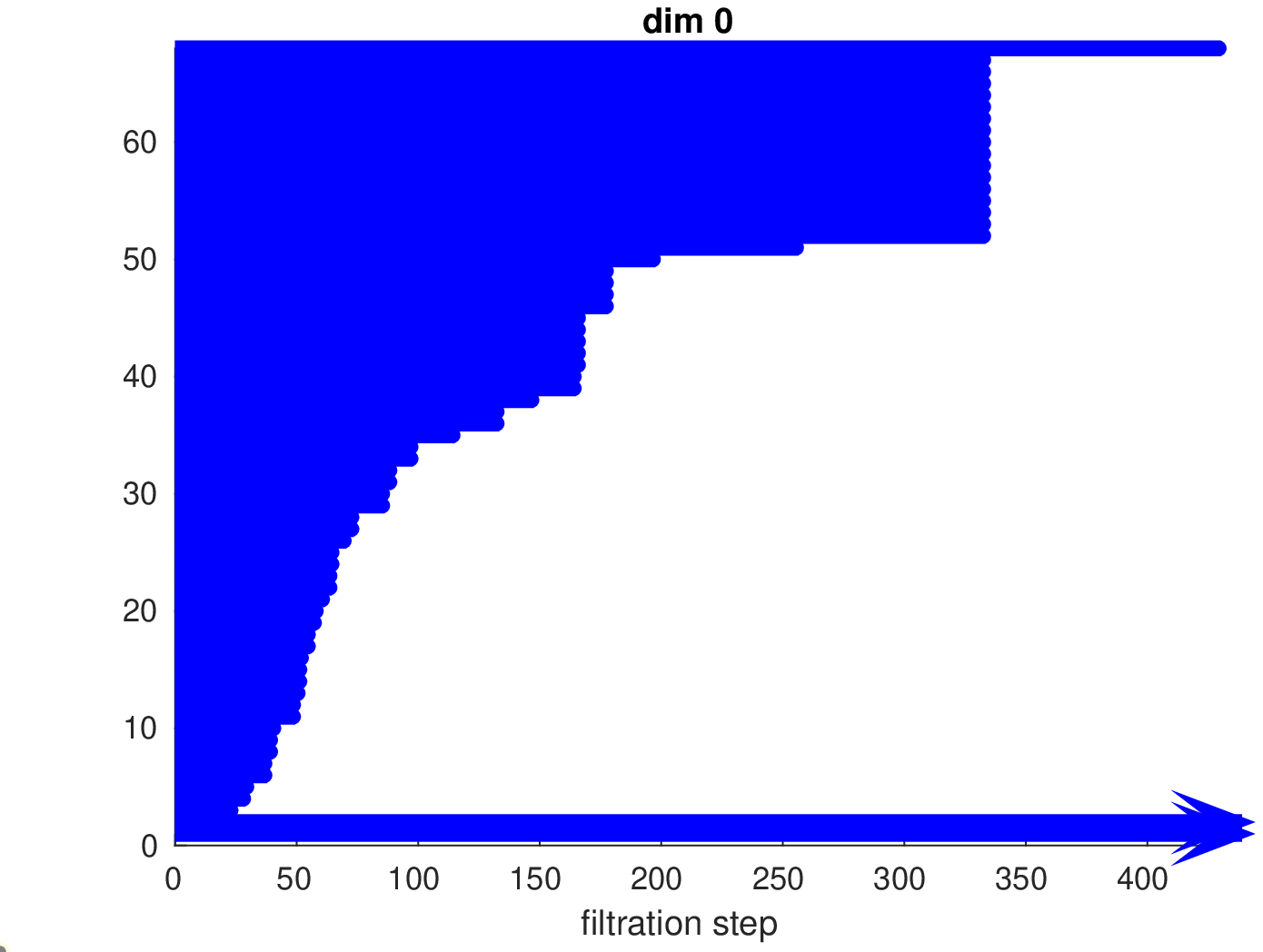}
  %\caption{A subfigure}
  %\label{fig:sub1}
\end{subfigure}%
\begin{subfigure}{.49\textwidth}
  \centering
  \includegraphics[width=1\linewidth]{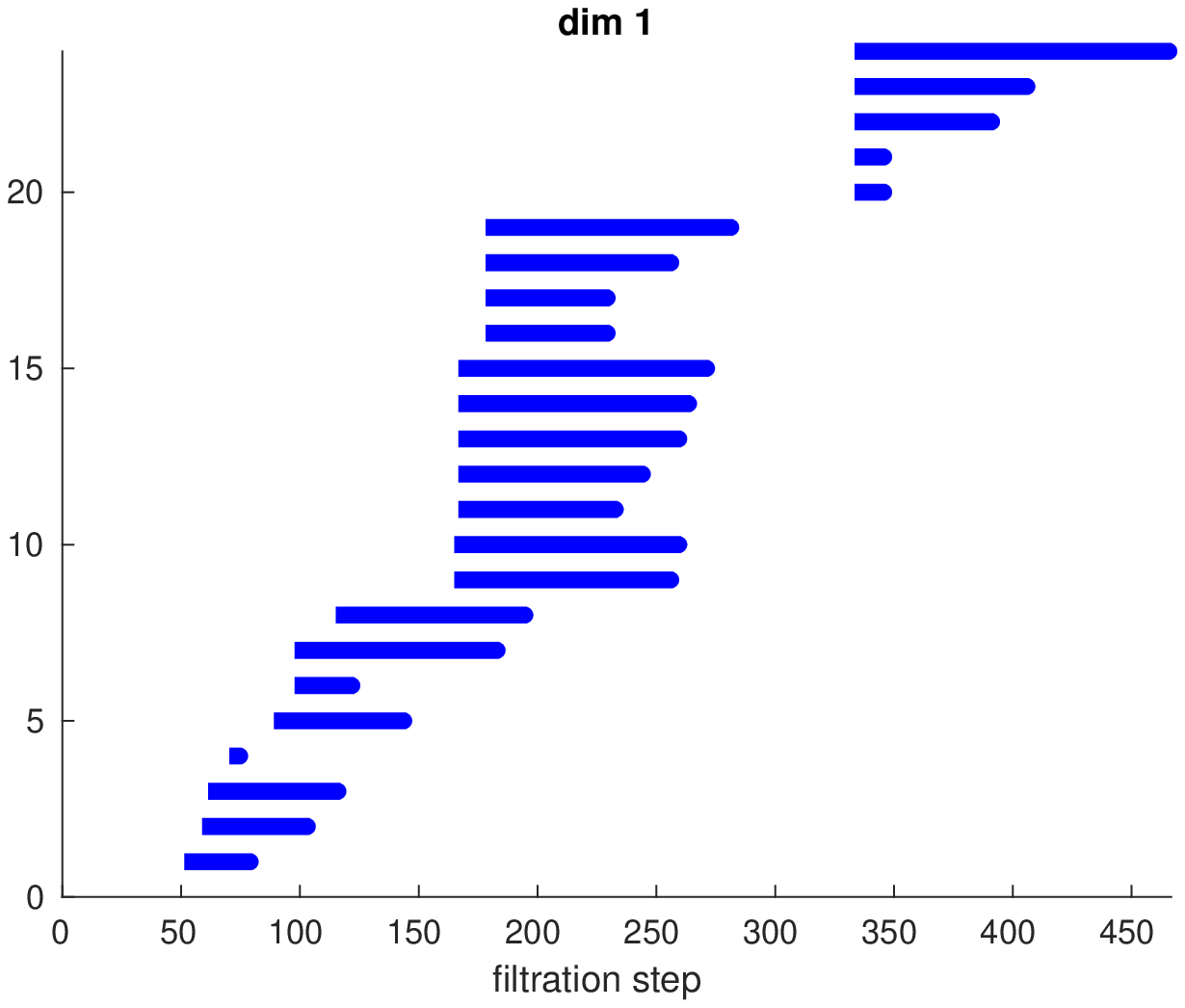}
  %\caption{A subfigure}
  %\label{fig:sub2}
\end{subfigure}
\caption{Barcodes of dimension 0 and 1 (left and right respectively) 
obtained from the first violin of the second movement of the String Quartet Op.17 by Haydn.}
\label{fig:bc}
\end{figure}
    \end{ex}

\newpage
\subsection{Statistical measures of information}

The main information in the persistent homology analysis is contained in the lengths of the bars,
which intuitively represent the importance of each corresponding feature.
For this reason we now employ some standard statistical measures of information in order to extract 
quantitative and qualitative features of the distributions of such lengths.
In particular, we will consider the mean, the variance and the entropy of the lengths, being
defined as follows (see also~\cite{rucco2016characterisation}).
    \begin{df}\rm\label{def:entropy}
    Consider a filtration $F$ given by  $K_0\subseteq K_1\subseteq\cdots\subseteq K_n$,   
    and the corresponding $p$-persistent barcode 
    $$ bc_p(F)=\{[x_i, y_i)|1\leq i\leq r,\,\, r\in\mathbb{N}\,\},$$ 
    where each interval corresponds to a bar starting at $x_i$ and ending at $y_i$. 
    Define the \emph{length of a bar} as $l_i=y_i-x_i$
    and let $I\subseteq\{0,1, ..., r\}$ be the subset of indices such that $y_i$ is finite.
%    $$ L=\{l_i=y_i-x_i|1\leq i \leq n\}.$$ 
    Then the \emph{p-persistent mean} and the \emph{p-persistent standard deviation} 
    are calculated according to their standard statistical definitions
    as follows:
    \begin{eqnarray}
    m_p(F)&=&\frac{1}{|I|}\sum_{i\in I}l_i\label{eq:mean}\\
    sd_p(F)&=&\sqrt{\frac{1}{|I|-1}\sum_{i\in I}\left(l_i-m_p(F)\right)^2}\label{eq:std}
    %e(F)&:=&-\sum_{i=1}^np_i\log(p_i)\label{eq:entropy}
    \end{eqnarray}
    Moreover, let $m=\max\{y_i|\,y_i\mbox{ is finite}\}$
    and set $y'_i=y_i$ if $y_i$ is finite and $y'_i=m+1$ otherwise. 
    Define $l'_i=y'_i-x_i$ and let $$p_i=\frac{l'_i}{\sum_{i=1}^r l'_i}$$
    be the distribution of the finite bar lengths. 
    Then the \emph{p-persistent entropy} is defined as
    \begin{eqnarray}
    e_p(F)&=&-\sum_{i=1}^rp_i\log(p_i)\label{eq:entropy}\,.
    \end{eqnarray}
    \end{df}
     
In the next section we will use these tools in order to assign to each musical piece 
a unique topological footprint that we will use in order to address questions such as
the stylistic exploration and variety of different authors with respect to a common genre
and the dual perspective of comparing the richness of various genres for a given author.

\section{Applications and results}\label{minuets}

\subsection{Same genre, different authors}
In this section we want to compare the musical style of different authors within a fixed genre
using the techniques described above.
To be precise, we consider the three  most representative string quartet composers
in classic style, 
namely Haydn, Mozart and Beethoven, and compare some of their string quartets. 
For each author we choose a set of representative works of each stage of his musical 
compositional development. 
We summarize these works in Tables~\ref{table:haydn}, \ref{table:mozart} and~\ref{table:beethoven}. 

\begin{table}[ht]
    \centering
    \begin{tabular}{ccccc}
    Publication & Date & Title & Key & Movement number \\
   \hline Op. 17/2 & 1771 & Menuetto & F major & 2 \\
    Op. 20/1 & 1772 & Menuetto & Eb major  & 2 \\
    Op. 33/3 & 1781 & Scherzando & C major & 2 \\
    Op. 50/3 & 1787 & Menuetto  & Eb major & 3\\
    Op. 64/1 & 1790 & Menuetto  & C major & 2 \\
    Op. 71/1 & 1793 & Menuetto  & Bb major & 3 \\
    Op. 77/1 & 1799 & Menuetto  & G major & 3\\
    \hline
\end{tabular}
    \caption{Haydn string quartets}
    \label{table:haydn}
\end{table}
\begin{table}[ht]
    \centering
    \begin{tabular}{ccccc}
    Publication & Date & Title & Key & Movement number \\
   \hline 
%   Quartet No. 1 & 1770 & Menuetto & G major & 3 \\
   Quartet No. 5 & 1773 & Tempo di Minuetto & F major  & 3 \\
   Quartet No. 8 & 1773 & Menuetto & F major  & 3 \\
   Quartet No. 13 & 1773 & Menuetto & D minor  & 3 \\
   Quartet No. 14 & 1782 & Menuetto & G major  & 2  \\
   Quartet No. 19 & 1785 & Menuetto & C major  & 3  \\
   Quartet No. 23 & 1790 & Menuetto-Allegretto & F major  & 3 \\
    \hline
\end{tabular}
    \caption{Mozart string quartets}
    \label{table:mozart}
\end{table}
\begin{table}[ht]
    \centering
    \begin{tabular}{ccccc}
    Publication & Date & Title & Key & Movement number \\
   \hline 
   Op. 18/1 & 1798 &  Scherzo: Allegro molto & F major & 3 \\
   Op. 18/4 & 1798 & Menuetto: Allegretto & C major  & 3 \\
   Op. 18/6 & 1798 & Scherzo: Allegro & Bb major  & 3 \\
   Op. 59/3& 1805 & Menuetto: Grazioso& C major  & 3 \\
   Op. 127& 1825 & Scherzando vivace - Presto & Eb major  & 3  \\
    \hline
\end{tabular}
    \caption{Beethoven string quartets}
    \label{table:beethoven}
\end{table}

Each of these works consists of four instruments (parts) and to each of them we may apply persistent homology as described above. 
For each musical work and each instrument we thus obtain the corresponding barcodes in dimension 0 and 1.
We then calculate the mean, standard deviation and entropy of their lengths according to Eqs.~\eqref{eq:mean},
\eqref{eq:std} and \eqref{eq:entropy}, thus obtaining $6$ statistical descriptors for each instrument,
for a total of $24$ descriptors for each work.
Then each work is described by a point 
    $$(m^1_0, sd^1_0, e^1_0, m^1_1, sd^1_1, e^1_1, m^2_0, sd^2_0,...,sd^4_1,e^4_1)\in \mathbb{R}^{24}$$ 
where $m^i_j$, $sd^i_j$ and $e^i_j$, with $i=1,\dots,4$ and $j=0,1$,  
stand for the mean, standard deviation and entropy of the barcodes in dimension $j$ of the $i$-th instrument, respectively.
Therefore for each composer we obtain a set of points in $\mathbb{R}^{24}$. 

\begin{figure}[ht]
    \centering
    \includegraphics[scale=0.5]{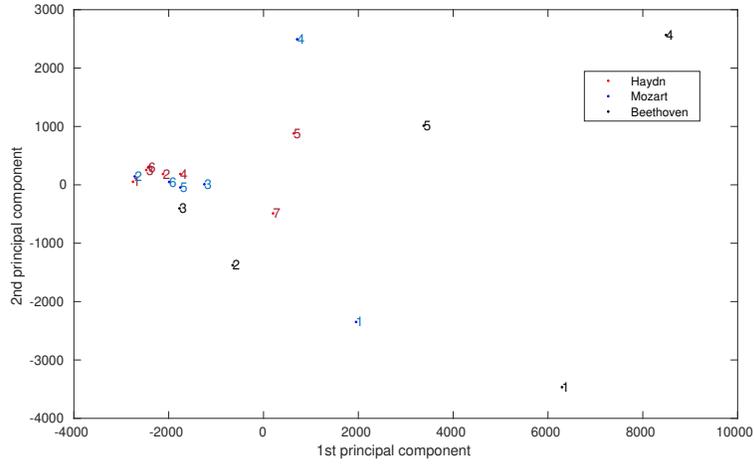}
    \caption{Principal component analysis of the considered works.
    Labels correspond to the chronological order in their composition.}
    \label{fig:pca}
\end{figure}

Finally, in order to reduce the dimensionality of the problem and obtain a succint but reliable visual
description of each work, 
we plot all these points in $\mathbb{R}^2$ using principal component analysis (PCA)~\cite{Jolliffe:1986}. 
In Figure~\ref{fig:pca} we illustrate the results of this analysis when considering only the first two principal components.

We now focus on the dispersion of the points for each author, meaning the 
deviation from their centre of mass. %mean position.

\begin{figure}[h!]
    \centering
    \includegraphics[scale=0.5]{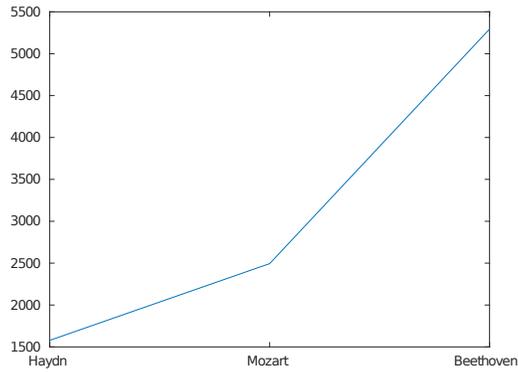}
    \caption{Dispersion among works by the same author as a quantitative indicator of their stylistic exploration
    withing the genre of string quartets.}
    \label{fig:disp}
\end{figure}

At first sight it seems clear from Fig.~\ref{fig:pca}
that the dispersion is 
increasing in the order Haydn, Mozart and Beethoven.
Indeed, we can make this statement quantitative by computing the dispersion as follows. 
    \begin{df}
    Let $x_1, x_2, ..., x_l$ be the set of points in $\mathbb{R}^{24}$ representing the works of a given composer 
and let $\bar{x}=\frac{1}{l}\sum_{i=1}^lx_i$. Then the \emph{dispersion} $\sigma$ is computed as
    $$ \sigma:=\sqrt{\frac{1}{l-1}\sum_{i=1}^l|x_i-\bar{x}|}\,.$$
    \end{df}
The results for each author are plotted in Figure~\ref{fig:disp}.
We can clearly observe that there is an increasing behavior when we move from Haydn to Mozart and from
Mozart to Beethoven.

This has an interesting musicological interpretation. 
Indeed, it is natural to associate the dispersion of the works considered as a quantitative measure 
of the stylistic variability. In other words, the dispersion provides an indicator of the scope of stylistic 
exploration by each of these composers. 
Therefore our result coincides with the standard view in which Haydn is seen as the initiator 
of the string quartet and with a compositional style that remained relatively stable during his life. Correspondingly, 
Mozart is perceived as an innovator, but probably due to his early death, his stylistic range did not change 
as much as it could have,
%\alescomment{it is correct (Pablo used a sophisticated but very nice to read construction)}
had he lived longer. Finally, the big variability shown by Beethoven's 
works coincides with the conception of this author as the most innovative, 
expanding and modifying the string quartet as a musical form~\cite{rosen1997classical}.

\subsection{Same author, different genres}
Contrary to the previous section, in this one 
we want to fix the author and compare works belonging to different genres.

Precisely, for every author (Haydn, Mozart and Beethoven) we have selected a set of works belonging to
different subgenres, such as minuets, allegros, adagios, etc. 
Namely, we chose the four most explored subgenres for each composer.
Then we applied the same analysis as in the previous section, obtaining for each work first a point in $\mathbb{R}^{24}$
and then the corresponding projection to $\R^2$ given by PCA.
The full list of the works considered for each author is available 
at \href{https://github.com/MartinMij/TDA-SQ}{https://github.com/MartinMij/TDA-SQ},
where the codes used can also be found.

\begin{figure}[ht]
    \centering
    \begin{subfigure}{.49\textwidth}
     \centering
     \includegraphics[width=1\linewidth]{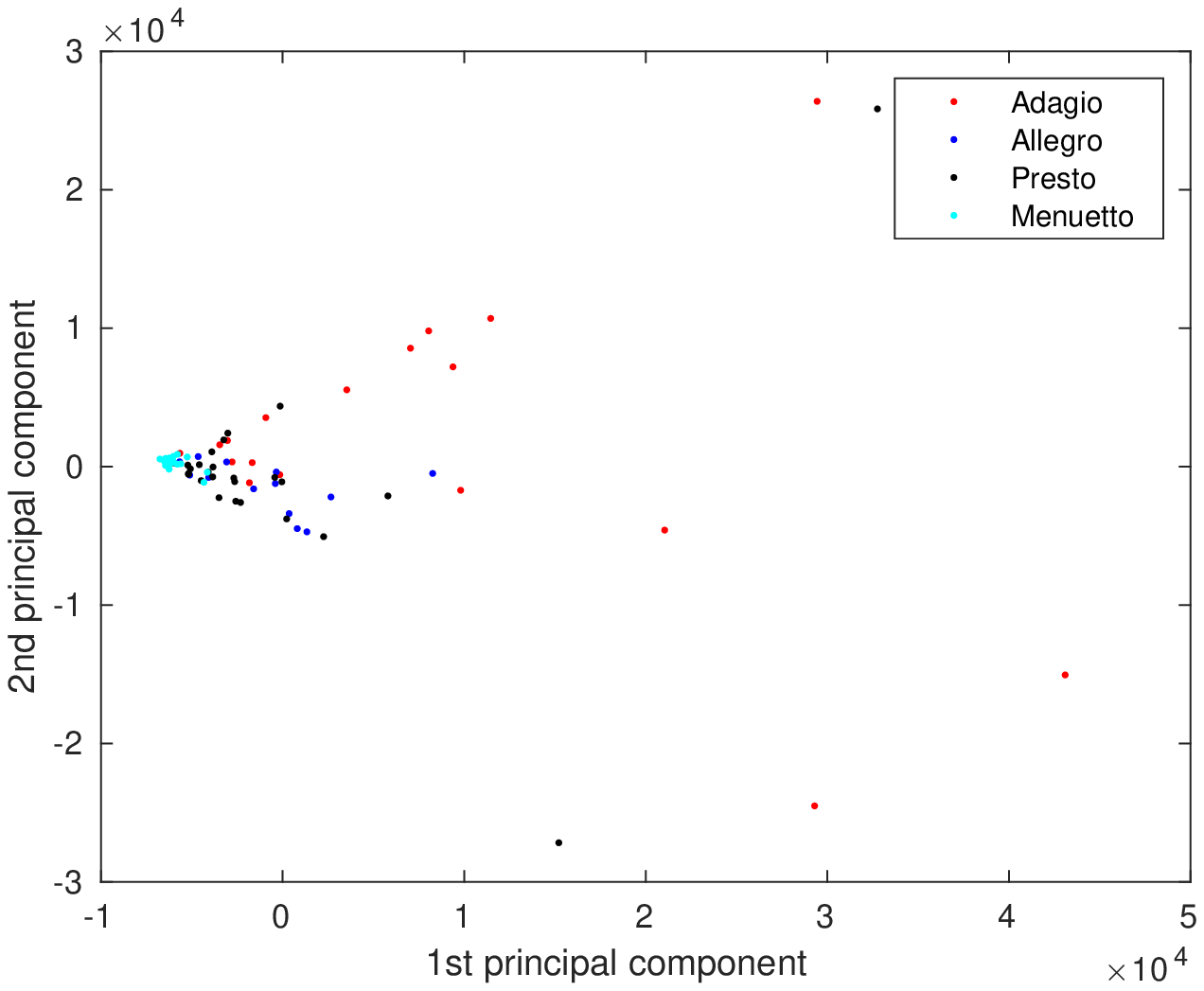}
    \end{subfigure}
    \begin{subfigure}{.49\textwidth}
      \centering
      \includegraphics[width=1\linewidth]{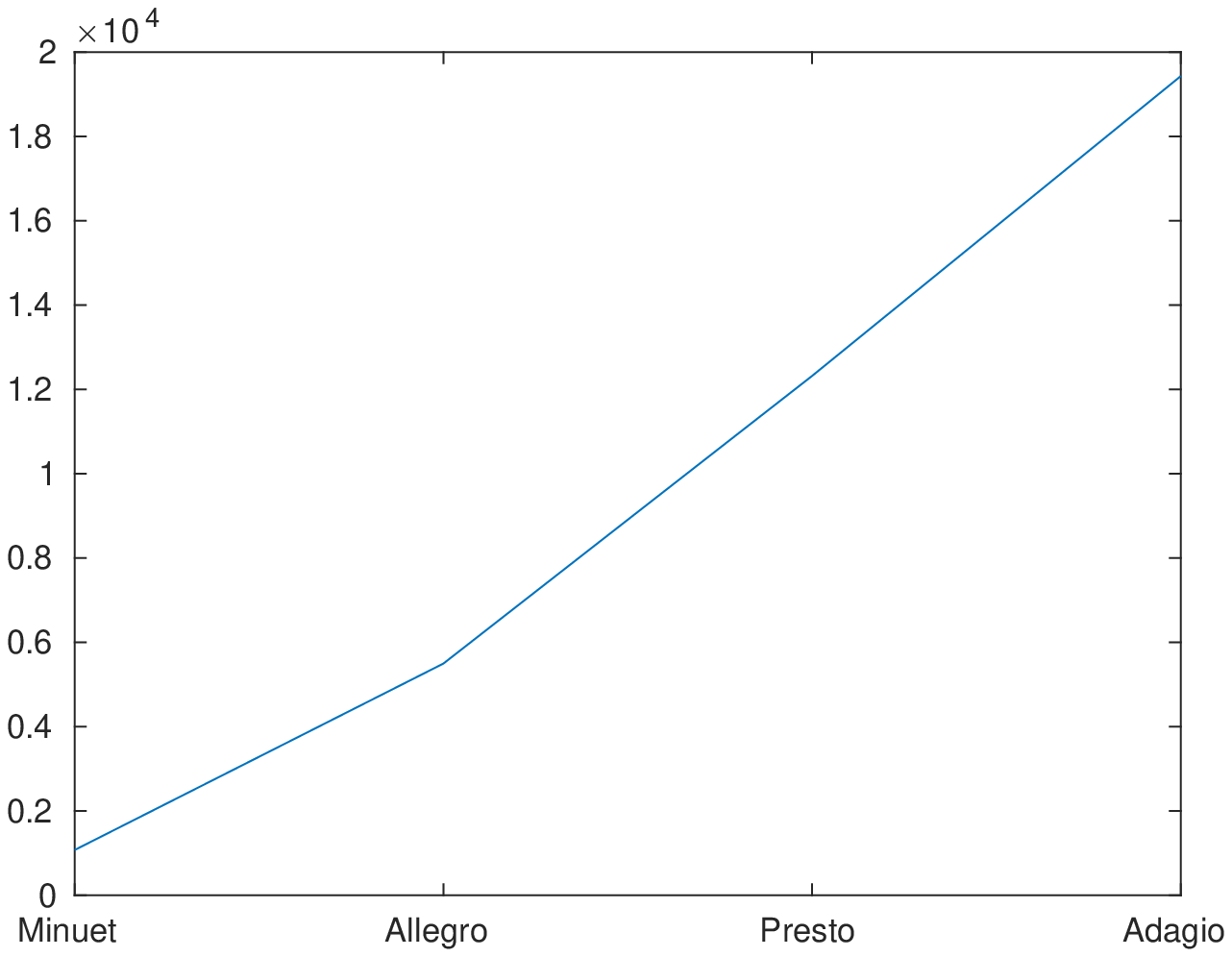}
    \end{subfigure}
    \caption{Analysis of several works of Haydn belonging to different subgenres. 
    In the left panel we display the results of the PCA analysis. In the right panel 
    we compute the dispersions of the points for each subgenre.}
    \label{subgenreh}
\end{figure}

\begin{figure}[ht]
    \centering
    \begin{subfigure}{.49\textwidth}
     \centering
     \includegraphics[width=1\linewidth]{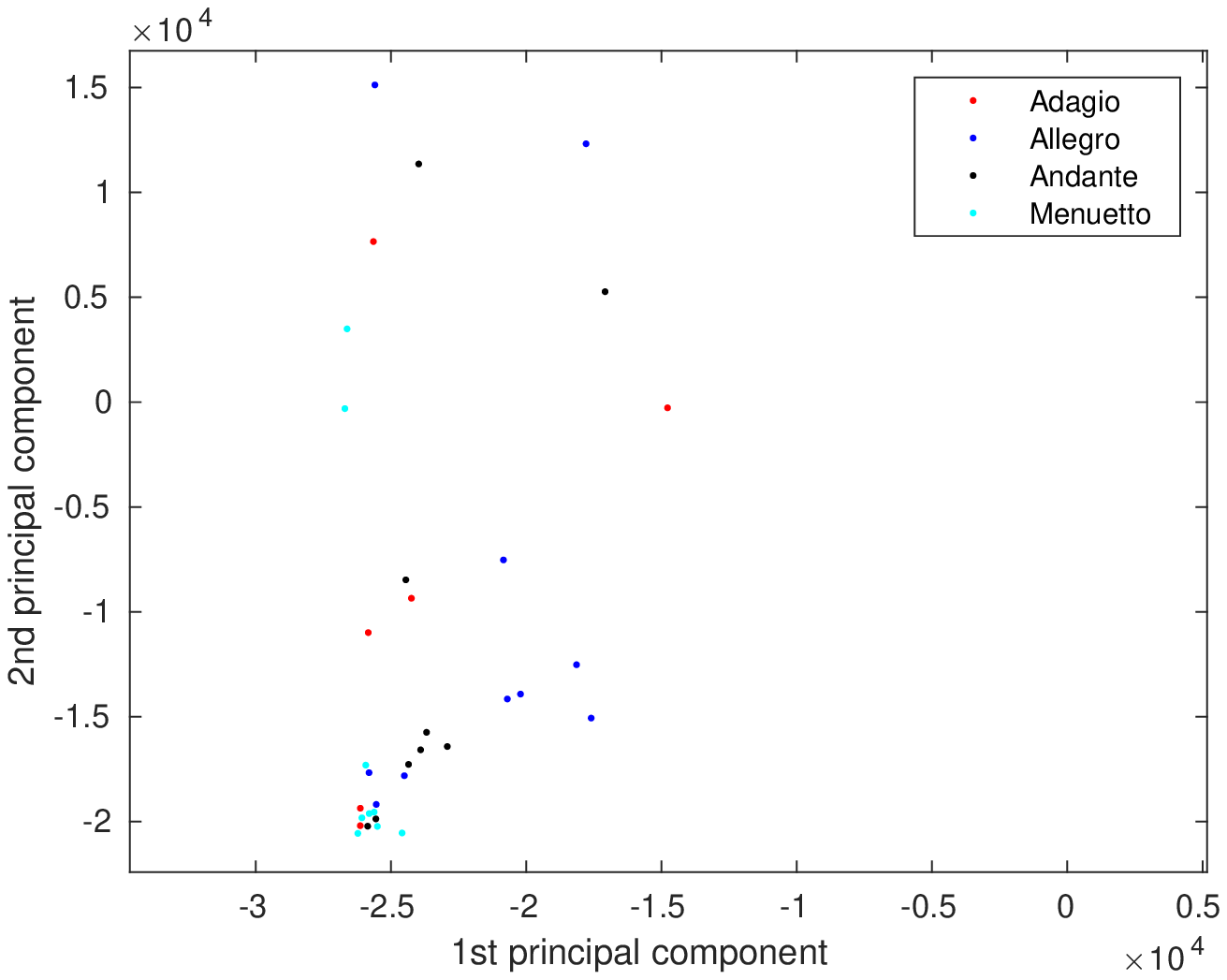}
    \end{subfigure}
    \begin{subfigure}{.49\textwidth}
      \centering
      \includegraphics[width=1\linewidth]{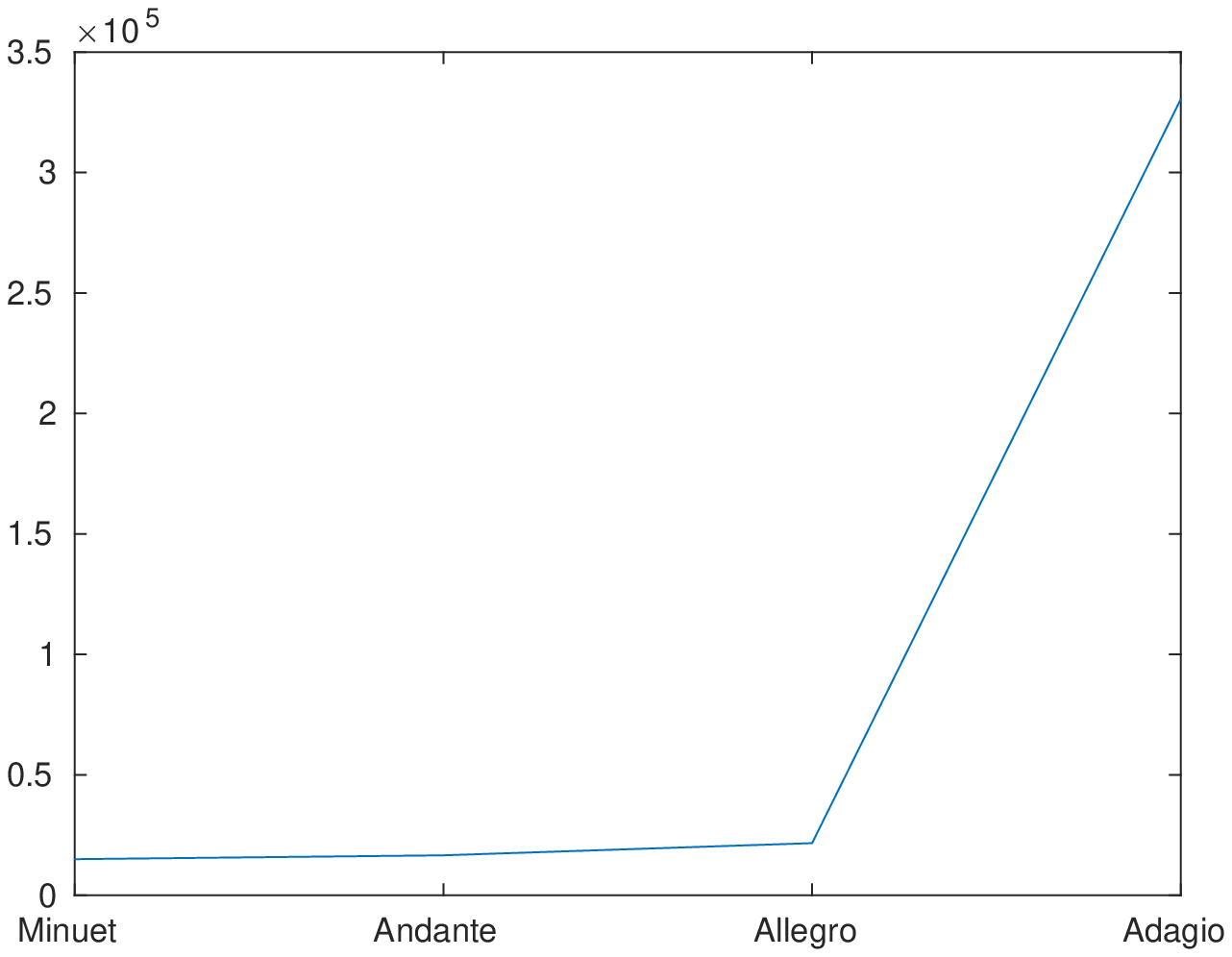}
    \end{subfigure}
  \caption{Analysis of several works of Mozart belonging to different subgenres. 
    In the left panel we display the results of the PCA analysis 
    (here two adagios that were outliers have been removed in order to better appreciate the dispersion). 
    In the right panel 
    we compute the dispersions of the points for each subgenre.}    
    \label{subgenrem}
\end{figure}

\begin{figure}[ht]
    \centering
    \begin{subfigure}{.49\textwidth}
     \centering
     \includegraphics[width=1\linewidth]{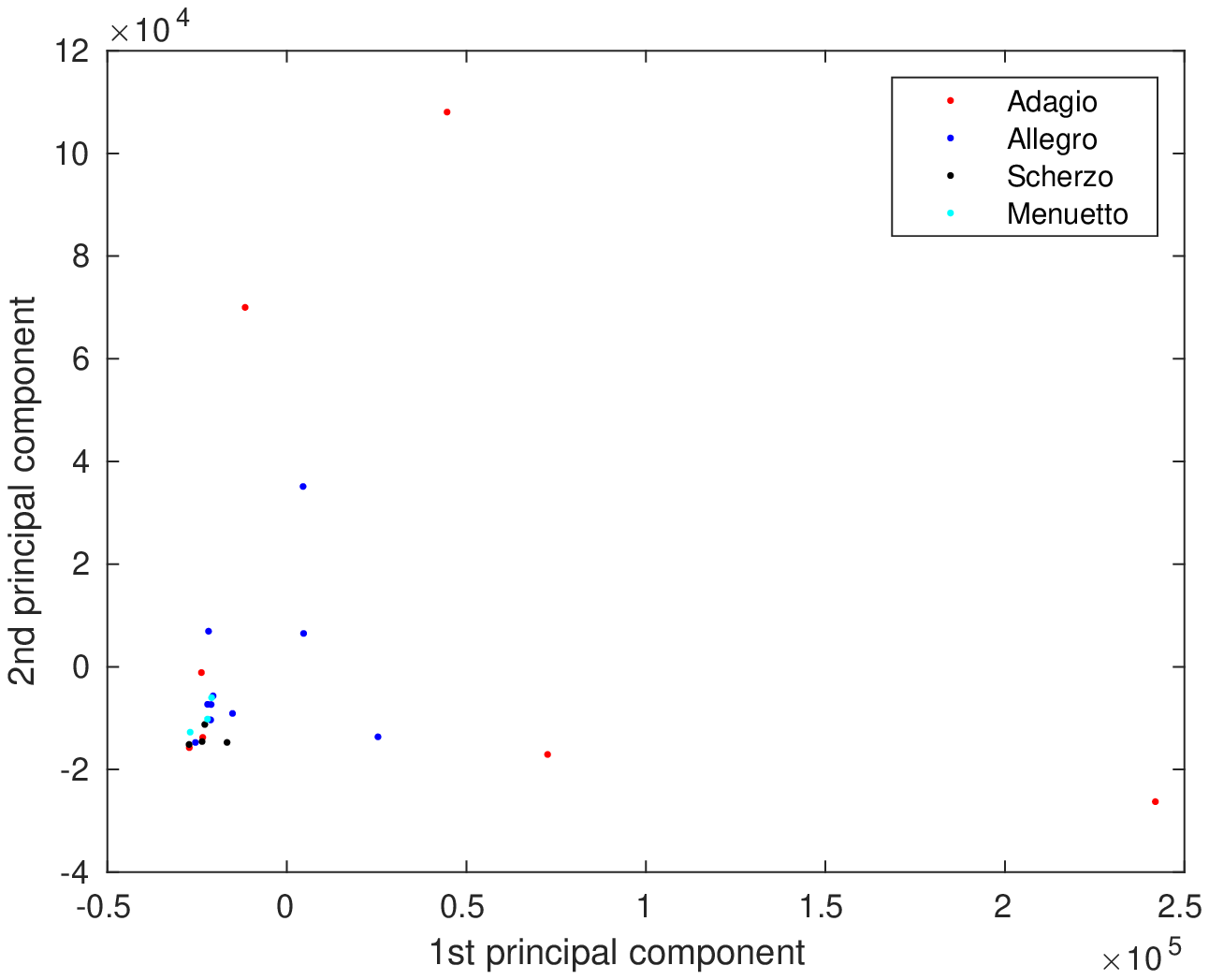}
    \end{subfigure}
    \begin{subfigure}{.49\textwidth}
      \centering
      \includegraphics[width=1\linewidth]{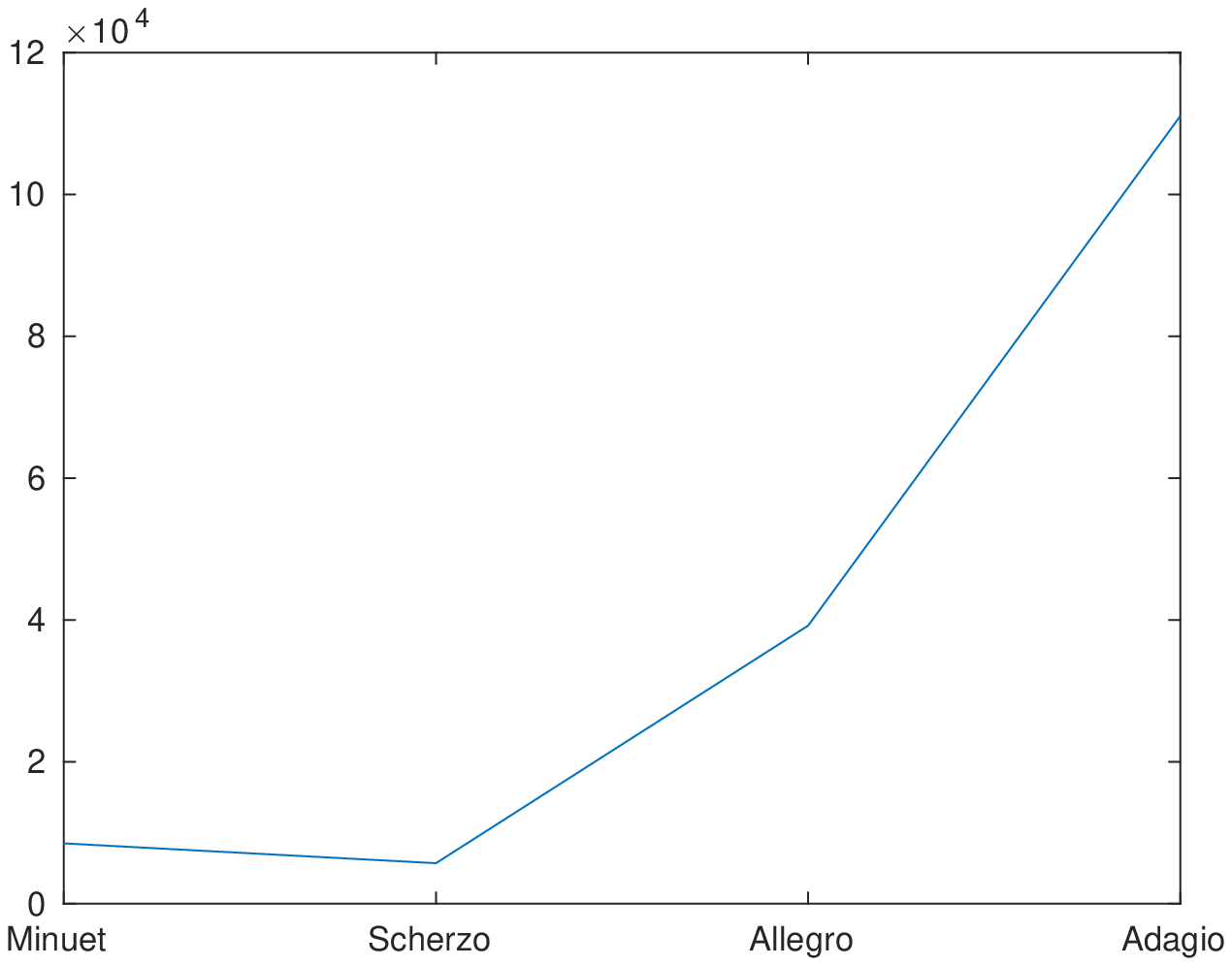}
    \end{subfigure}
  \caption{Analysis of several works of Beethoven belonging to different subgenres. 
    In the left panel we display the results of the PCA analysis. In the right panel 
    we compute the dispersions of the points for each subgenre.}    
    \label{subgenreb}
\end{figure}

The results thus obtained are shown in the 
left panels of Figures~\ref{subgenreh}, \ref{subgenrem} and~\ref{subgenreb},
while 
in the right panels we display the corresponding dispersions for each author.

As we can see, 
the dispersion of the different types of subgenre changes with each composer, 
but there are some interesting regularities:
on the one hand, the minuets are in general the subgenre with the least dispersion, confirming 
the fact that the minuet has a more uniform formal and stylistic structure.  
On the other hand, 
the adagios have the greatest dispersion for all the authors, indicating that this subgenre
is the most versatile in terms of stylistic exploration.

%\newpage

%\subsection{A glimpse into the rests}
\subsection{What if we take rests?}
Hitherto we have considered the pitch class distribution without taking into account the rests of a given score. 
A natural question is how much our previous results change if we either 
do not consider transitions between tones that are separated by a rest, 
or if we assign to them a smaller weight. 

To test the robustness of our results under such changes, we modified 
the pitch class distribution in two ways:
in the first case we removed completely the transitions that contain a rest, while in the second case
we assigned to them a weight according to the following procedure:
suppose we have a transition $(i, j)$ (i.e.~from the pitch class $i$ to $j$) where the first tone has a duration $u$ 
and the second one has a duration $v$, separated by a silence of duration $s$. 
Instead of considering $uv$ as the total duration of the transition $(i, j)$, as it was the case so far,
now we assign to this transition the value 
$uvf$ where $0\leq f \leq 1$ 
is a function of $u$, $v$ and $s$ that weights $uv$ so that transitions 
with long silences between shorts tones add little to the distribution, 
while transitions with short silences 
between long tones are more important for the distribution. 
Specifically, we take $f$ to be
$$f=\dfrac{1}{\frac{s}{uv}+1}\,.$$
Note in particular that for $s=0$ we obtain $f=1$ and the duration of the transition is unchanged, as it should be.

\begin{figure}[ht]
\centering
\begin{subfigure}{.3\textwidth}
  \centering
  \includegraphics[width=0.9\linewidth]{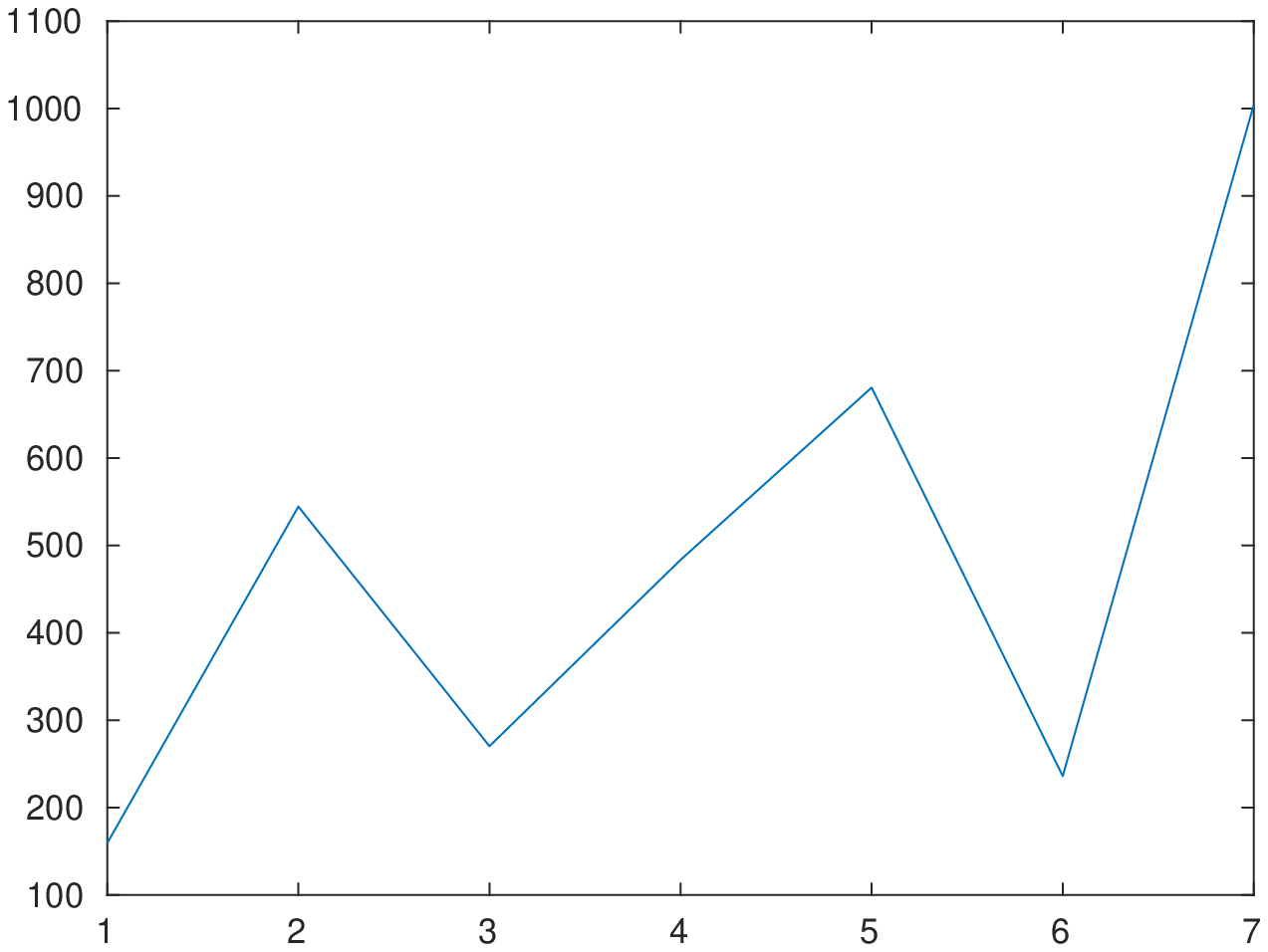}
  \caption{Original}
  %\label{fig:sub1}
\end{subfigure}%
\begin{subfigure}{.3\textwidth}
  \centering
  \includegraphics[width=0.9\linewidth]{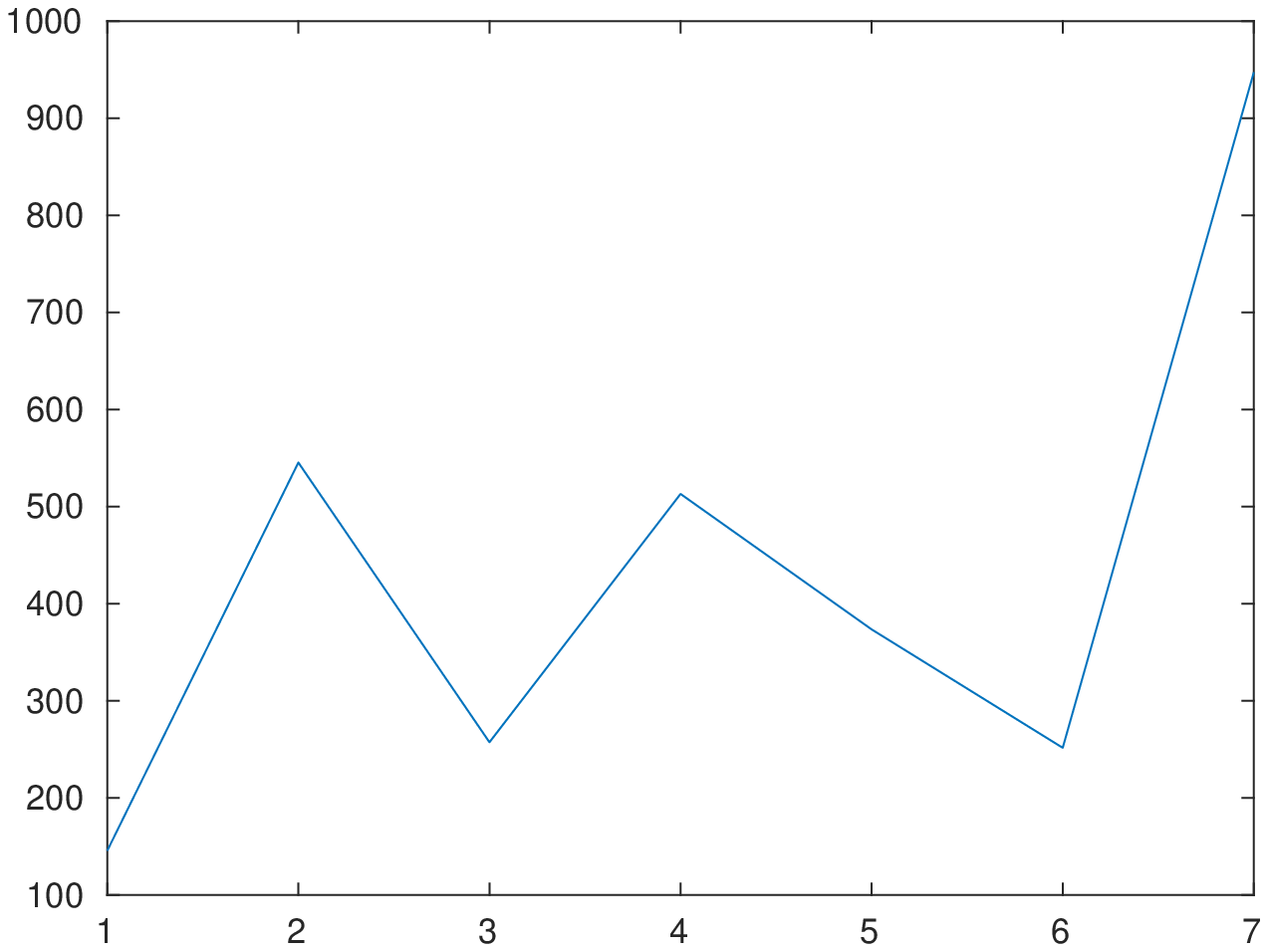}
  \caption{Rests considered}
  %\label{fig:sub2}
\end{subfigure}
\begin{subfigure}{.3\textwidth}
  \centering
  \includegraphics[width=0.9\linewidth]{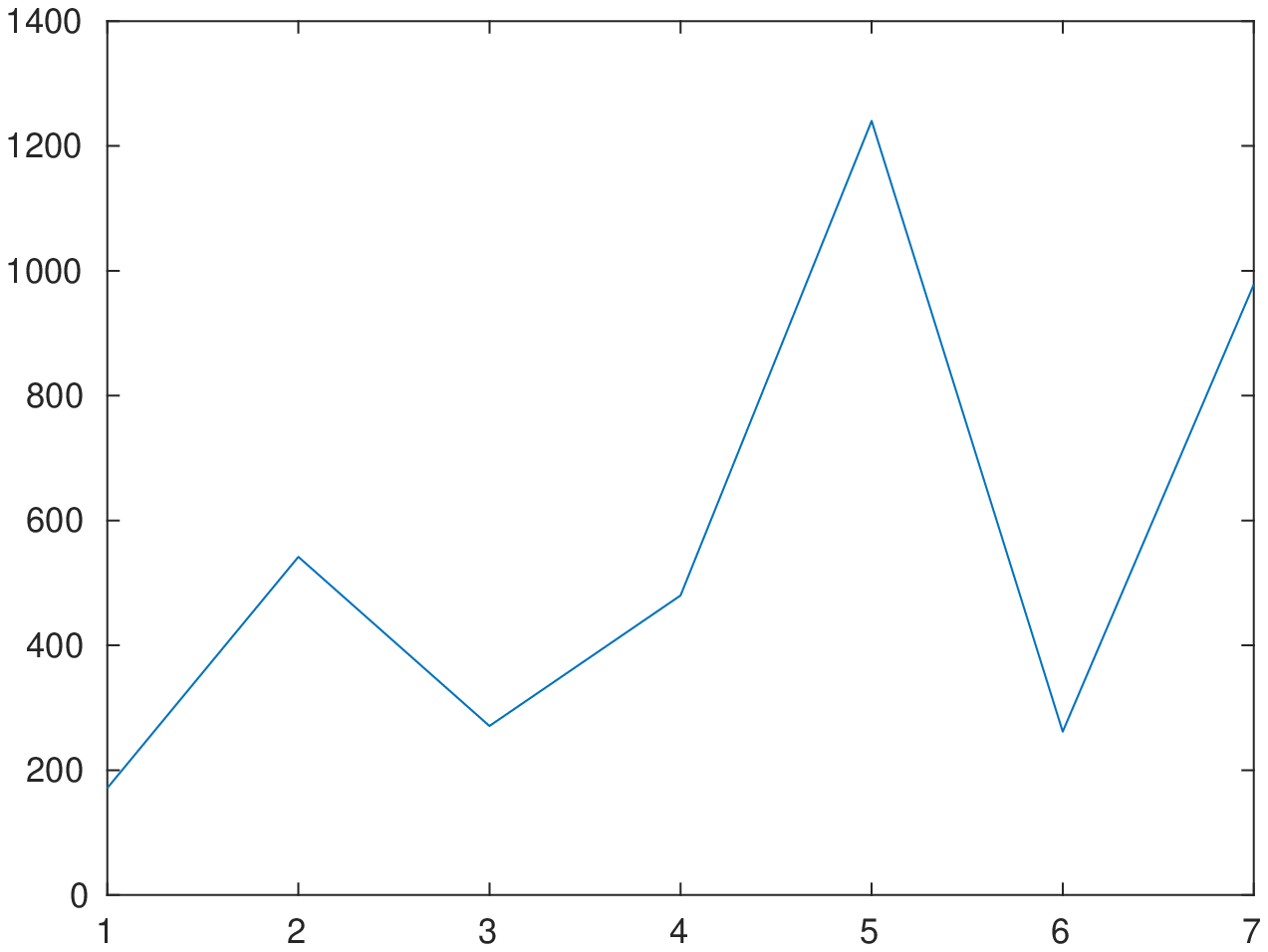}
  \caption{Weighting by f}
  %\label{fig:sub2}
\end{subfigure}
\caption{0-persistence mean considering three versions of the pitch class distribution. 
In (a) the transitions are taken as described in Subsection~\ref{phofmusicalworks}. 
In (b) transitions with a rest in the middle are omitted. 
In (c) the transitions are weighted by $f$.}
\label{pcd}
\end{figure}

\begin{figure}[h]
    \centering
    \includegraphics[scale=0.3]{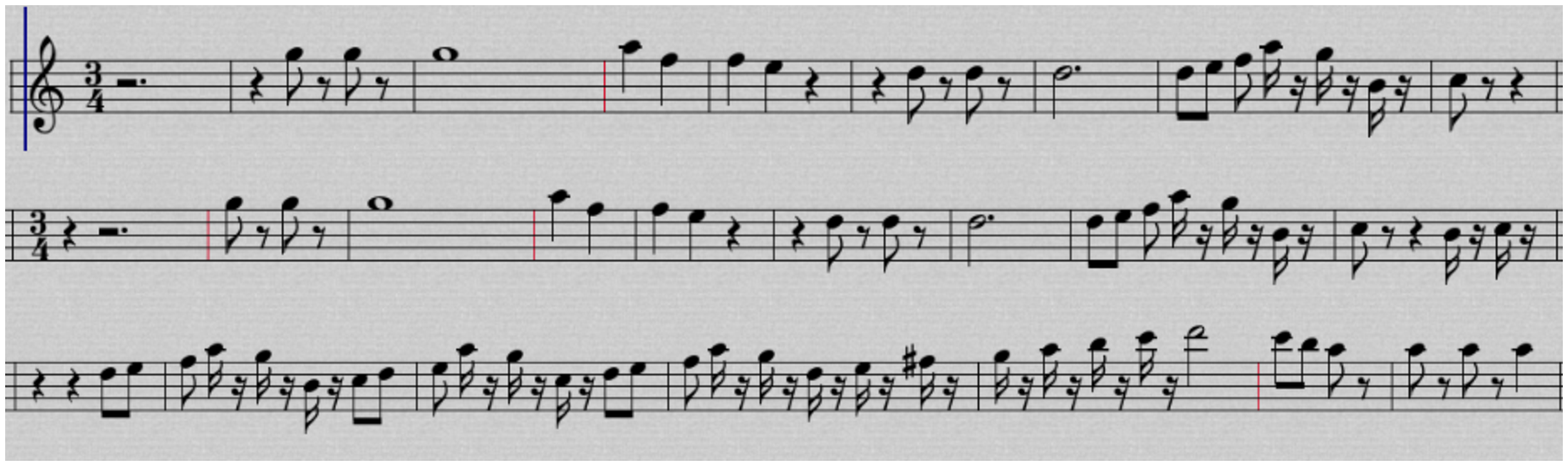}
    \caption{First measures of the String Quartet Op. 64/1 by Haydn.}
    \label{5th}
\end{figure}

For the purpose of comparison, we apply persistent homology to the first violin of Haydn's 
string quartets minuets shown in Table \ref{table:haydn}.  
As an illustration, in Figure~\ref{pcd} we show
the 0-persistence means  for each of the seven works.
%\alescomment{these figures are not aligned. ``Mean'' or ``persistent mean''?}
As we can see, only the 5th work exhibits a major change. 
If we look at the score (in Figure~\ref{5th} we show the first measures) 
we realize why this happens. Indeed, many of the rests are very short, namely, sixteenth note rests, 
and are more to explicitly notate a desired articulation, rather than structural rests. 

\section{Conclusions and future work}\label{conclusions}
We have developed a method to obtain quantitative measures to compare composition styles.
Specifically, this method takes into account the duration of the notes involved in the played transitions of the work 
and results in a metric space where closer points correspond 
to more related pitch class tones. Persistent homology allows us to summarize this information in barcodes 
and through statistics and information-theoretical
tools, we assign a point in a Euclidean space to each musical work. 
Then we can see how stylistically different two works are by comparing their associated points. 

Our results from the analysis of the string quartets of Haydn, Mozart and Beethoven
are consistent with the standard view of Haydn as the initiator of the genre and Beethoven as the most innovative
author among the three. Moreover, we were also able to make quantitative statements about the stylistic variety among
different genres by the same author, and found out that minuets are the most uniform structure, 
while adagios are the most versatile. Finally, the results are robust with respect to different ways to consider rests
in the transitions.

In future research we will consider two important aspects. 
First, while the methods developed here 
%we have applied our analysis to a specific corpus of string quartets,}
apply 
to any set of works as long as all of them use the same 
 instruments, 
it will be interesting to extend our techniques and adapt them to works having different instruments.
%this with the aim that all works fit in the same Euclidean space.
As a second point, our method depends on the use of a metric space. 
The distance  is obtained from a weighting function on the edges of a graph. 
If the graph is directed, the weighting function is not symmetric in general and neither is the distance. 
This is the reason why we had to induce an undirected graph from a directed one. But what if we could work 
with a non-symmetric ``distance" and still be able to obtain filtrations and then compute persistent homology?  
A perspective in this direction is given in~\cite{edelsbrunner2016topological}. 
We expect to develop this approach in future work.

%------------------------------------------------------------------------------------------------------------------------------------------------------
%------------------------------------------------------------------------------------------------------------------------------------------------------
%------------------------------------------------------------------------------------------------------------------------------------------------------

%\bibliographystyle{abbrvnat_mv}
\bibliographystyle{abbrv}
\bibliography{references.bib}

%------------------------------------------------------------------------------------------------------------------------------------------------------
%------------------------------------------------------------------------------------------------------------------------------------------------------
%------------------------------------------------------------------------------------------------------------------------------------------------------

\end{document}